\documentclass[aps,prd,nopacs,nofootinbib,notitlepage]{revtex4-1}

\usepackage{amsfonts}
\usepackage{amsmath}
\usepackage{xcolor}
\usepackage{bm}
\usepackage{graphicx}
\usepackage{fancybox}
\usepackage{comment}
\usepackage{multirow}
\usepackage{tipa}
\usepackage{longtable}
\definecolor{refs}{RGB}{245,156,74}
\usepackage[colorlinks=true,hyperfootnotes=true,citecolor=cyan]{hyperref}

\usepackage{enumitem}

\newcommand{\be}{\begin{equation}}
\newcommand{\ee}{\end{equation}}
\newcommand{\ba}{\begin{eqnarray}}
\newcommand{\ea}{\end{eqnarray}}
\newcommand{\bs}{\begin{subequations}}
\newcommand{\es}{\end{subequations}}

\newcommand{\diff}{\textrm{d}}
\newcommand{\Diff}{\textrm{D}}

\newcommand{\lp}{\left(}
\newcommand{\rp}{\right)}
\newcommand{\lb}{\left[}
\newcommand{\rb}{\right]}

\newcommand{\mJ}{\mathcal{J}}
\newcommand{\mX}{\mathcal{X}}
\newcommand{\mY}{\mathcal{Y}}

\newcommand{\nn}{\nonumber}

\newcommand{\tomi}[1]{{\textcolor{red}{ #1}}}

\begin{document}

\title{Relativistic interacting fluids in cosmology}

\author{Damianos Iosifidis}
 \email{damianos.iosifidis@ut.ee}
\address{Laboratory of Theoretical Physics, Institute of Physics, University of Tartu, W. Ostwaldi 1, 50411 Tartu, Estonia.}

\author{Erik Jensko}
\email{erik.jensko@ucl.ac.uk}
\address{Department of Mathematics, University College London, \\ Gower Street, London WC1E 6BT, UK}

\author{Tomi S. Koivisto}
\email{tomi.koivisto@ut.ee}
\address{Laboratory of Theoretical Physics, Institute of Physics, University of Tartu, W. Ostwaldi 1, 50411 Tartu, Estonia}
\address{National Institute of Chemical Physics and Biophysics, R\"avala pst. 10, 10143 Tallinn, Estonia}

\begin{abstract}
Motivated by cosmological applications for interacting matters, an extension 
of the action functional for relativistic fluids is proposed to incorporate the
physics of non-adiabatic processes and chemical reactions. The former
are characterised by entropy growth, while the latter violate
particle number conservation. 
The relevance of these physics is demonstrated in the contexts of
self-interacting fluids, fluids interacting with scalar fields, and
hyperhydrodynamical interactions with geometry. The possible cosmological 
applications range from early-universe phase transitions to astrophysical
phenomena, and from matter creation inflationary alternatives to interacting
dark sector alternatives to the $\Lambda$CDM model that aim to address its tensions.
As an example of the latter, a single fluid model of a unified dark sector is presented. The simple action of the model features one field and one parameter, yet it can both reproduce the $\Lambda$CDM cosmology and predict new phenomenology.
\end{abstract}

\maketitle

\section{Introduction}


Scalar fields are ubiquitous in theoretical cosmology, yet there is no 
experimental evidence for the fifth forces they would mediate 
\cite{Wetterich:1987fm,Brax:2007ak,Burrage:2017qrf,Vagnozzi:2021quy,Fischer:2024eic}.
On the other hand, the case is quite the opposite with non-adiabatic and chemical reactions, which take place all around (and inside) us. Nonetheless, cosmological 
models are seldom systematically developed from the general thermodynamic perspective in terms of variations in entropy and in an effective particle number. 

In standard inflationary cosmologies, there are whole series of events in the very early universe (pre- and reheatings, baryo- and leptogeneses, etc.)  which call for various particle number violations that can be described by well elaborated, and, at least in the case of nucleosynthesis, well established, particle physics theories \cite{Linde:1990flp,Davidson:2008bu,Allahverdi:2010xz,Marzola:2015dia,Karam:2020skk}. In late-universe cosmology, and particularly in the context of the dark energy problem, a plethora of models have been proposed that introduce non-minimal interactions between dark energy and (dark or visible) matter \cite{Farrar:2003uw,Pourtsidou:2013nha,Wang:2016lxa,CarrilloGonzalez:2017cll,Teixeira:2024wsw}, featuring 
conformal \cite{Amendola:1999er,Koivisto:2005nr,Wetterich:2007kr,Kase:2020hst,Poulot:2024sex}, disformal \cite{Zumalacarregui:2012us,vandeBruck:2015ida,Cembranos:2016jun,Teixeira:2019hil,Vagnozzi:2019kvw,vandeBruck:2020fjo},
3-form \cite{Ngampitipan:2011se,Koivisto:2012xm,Barreiro:2016aln,Yao:2017enb,Yao:2020hkw}, geometric \cite{Koivisto:2005yk,Harko:2020ibn,Lobo:2022aop,Boehmer:2024rqk,Iosifidis:2024bsq}, elastic \cite{Asghari:2019qld,Figueruelo:2021elm,BeltranJimenez:2021wbq,Poulin:2022sgp,Cardona:2022mdq} or merely parameterised \cite{Valiviita:2008iv,Clemson:2011an,Yang:2017zjs,Yang:2018euj,deCesare:2021wmk,Ferlito:2022mok,Nunes:2022bhn} couplings. Currently, a strong motivation to consider such interactions is their potential to alleviate the existing tensions in the cosmological data, especially the Hubble tension \cite{Odderskov:2015fba,DiValentino:2019ffd,DiValentino:2019jae,Gomez-Valent:2020mqn,Hoerning:2023hks,Pan:2023mie}. 

{The aim of this article is to develop a universal and physically sound framework for the description of interacting fluids. 
Phenomenological models without a proper Lagrangian formulation are compromised due to their arbitrariness or lack of predictivity, as well as unphysical features like instabilities they can introduce (e.g. \cite{Valiviita:2008iv,Herrera:2016uci,Yang:2017zjs,Aljaf:2019ilr,deCesare:2021wmk}). 
Even in the case that the model of the dark sector is based on an elementary
particle theory}\footnote{{An interesting current example is the scenario wherein a fraction of QCD 
axion dark matter is converted by a novel level crossing mechanism into  
dark energy, which is provided by a pseudo-Nambu Goldstone boson exploiting
the temperature-scaling of its mass in a confining phase transition \cite{Muursepp:2024mbb,Muursepp:2024kcg}.}}, the collective description of the microscopic degrees of freedom at the level of a hydrodynamical Lagrangian can be useful in order to understand the implications of the model to cosmological large scale structures. As will be shown in this article, an opposite proceeding is also possible, to wit, a consistent quantum field theory can emerge from a phenomenological model of thermodynamical processes by integrating out variables from the relativistic fluid action functional. 

We shall propose the action formulation of interacting relativistic fluids, and utilise it to present several completely new classes of cosmological models. 
A virtue of our formulation is that it involves only the fundamental hydro- and thermodynamic fields, and these retain their physical interpretation despite the presence of the new couplings. {Interactions of the chemical type are incorporated via chemical momentum and non-adiabatic interactions via
thermasy. Even though the concept of thermasy does not often appear in the literature\footnote{See, however, an insightful discussion at \url{https://math.ucr.edu/home/baez/week289.html}.}, thermasy is one of the most fundamental quantities in thermodynamics \cite{VANDANTZIG1939673} as it spans the phase space of a system as the symplectic conjugate of entropy \cite{Kijowski:1979dj,Bao:1985hj}. 
On the other hand, chemical momentum is the symplectic conjugate of particle number density. Chemical reactions change the number of particles, for example, by combining molecules into new different molecules. The energy of such a process is proportional to the time variation of the chemical momentum: the chemical potential.} 
In Table \ref{symplectic} we list some of these analogies between different physical systems, including those that will be used throughout this work.

A general action functional incorporating the two types of interactions into a hydrodynamical description of matter is proposed in Section \ref{formalism}. We do not review the basic hydro- and thermodynamics in detail, but rather just list our definitions.  The formulation follows Ref. \cite{Misner:1973prb} and in particular Ref. \cite{Brown:1992kc}, though our identifications of the thermodynamic potentials are somewhat different and, in our opinion, more conventional. In Section \ref{scalar-fluid} we consider scalar field models of interacting matter in the new framework, and in Section \ref{hyperhydro} we consider interactions of matter and geometry; in particular, we propose an improved realisation of the so-called \textit{hyperhydrodynamics}, the description of fluids with non-minimal couplings to the affine connection. Section \ref{cosmo} discusses cosmological applications. We take one particular model as an example: a gas of self-interacting particles. The simple example turns out to provide a viable description of the cosmology's dark sector, the single matter component thus playing the double role of both dark matter and dark energy.  
We conclude in Section \ref{conclu}, briefly discussing future directions and applications.

\begin{center}
\begin{table}[h]
\begin{tabular}{  |c|c|c|c| c| } 
\hline
 variable & conjugate & $\diff$(conjugate)/$\diff t$ & $\Delta$variable$\times$$\diff$(conjugate)/$\diff t$  \\ 
 \hline
 position & momentum & force & mechanical moving energy \\ 
 orientation & angular momentum & torque & mechanical turning energy \\
 volume & pressure momentum & pressure & hydraulic energy \\
 charge & flux linkage & voltage & electric energy \\ 
 entropy & thermasy & temperature & thermal energy \\
 particle number & chemical momentum & chemical potential & chemical energy \\
 \hline
 magnitude & dilation & homothetic force & calibration energy \\
 shape & shear & stress & disformation energy \\
 \hline
 \end{tabular}
 \caption{\label{symplectic} Examples of the universal structure of symplectic conjugates in different areas of physics. Our formalism concerns the last two examples of the middle rows; more precisely, we will work in terms of the entropy per particle $s$ (aka the specific entropy) and obtain the temperature $T=\dot{\theta}$ as the derivative of thermasy $\theta$. Likewise, we consider the particle number density $n$, and find that its conjugate $\varphi$ is the chemical momentum, corresponding to the chemical potential $\mu = \dot{\varphi}$. Heat is a form of thermal energy and chemical energy is known as the Gibbs free energy in thermodynamics. Below the line, we have included heuristics of the hypermomenta considered in Section \ref{hyperhydro}.}
\end{table}
\end{center}

\section{Hydrodynamical interactions}
\label{formalism}

Following  \cite{Brown:1992kc,Boehmer:2015kta,Boehmer:2015sha}, a basic field in our formulation is the vector density $\mJ^\mu$, interpreted as the particle number flux density. The number density $n  =  \lvert \mJ\rvert/\sqrt{-g}$ is not an independent variational degree of freedom in this formulation. We can write $\mJ^\mu = \sqrt{-g}n u^\mu$, and so have at hand a covariant 1+3 decomposition in terms of the fluid 4-velocity $u^\mu$. This allows us to use a notation for dot as the time derivative defined by the fluid flow and $h_{\mu\nu} = g_{\mu\nu} + u_{\mu}u_{\nu}$ for a spatial metric. 
By considering an equation state $\rho = \rho(n,s)$, we are led to the following set of (implicitly) defined variables\footnote{We trust that the enthalpy per particle $h$ will not be confused with the number $3=h^\mu{}_\mu$ (nor the determinant of $h_{\mu\nu}$, which will not be needed in this paper).} 
\bs
\label{fluid}
\ba
\text{energy density} & : & \quad  \rho  = \rho(n,s)\,, \label{energydensity} \\
\text{4-velocity}  & : & \quad u^\mu  =  \mJ^\mu/\lvert \mJ \rvert \quad \text{where} \quad \lvert \mJ \rvert = \sqrt{-g_{\mu\nu}\mJ^\mu \mJ^\nu}\,, \\
\text{number density} & : & \quad  n = \lvert \mathcal{J}\rvert/\sqrt{-g}\,, \\
\text{entropy per particle} & : & \quad  s\,, \\
\text{pressure}  & : & \quad  p  =  n\frac{\partial \rho}{\partial n} - \rho\,, \label{pressure}  \\
\text{enthalpy per particle} & : &  \quad h  =  \frac{\partial \rho}{\partial n} = \frac{1}{n}\lp \rho + p\rp\,, \\
\text{chemical potential}  & : &  \quad \mu  =  h - Ts\,, \label{free} \\
\text{temperature} & : &  \quad T  =  \frac{1}{n}\frac{\partial\rho}{\partial s}\,.
\ea
\es
These definitions agree with\footnote{{In the notation and terminology of Misner, Thorne \& Wheeler \cite{Misner:1973prb} inherited by Brown \cite{Brown:1992kc}, our $h$ is denoted $\mu$ and called the chemical potential and our $\mu$ is denoted $f$ and called the chemical free energy. We rather use the conventions of standard textbooks on thermodynamics. In Ref. \cite{Misner:1973prb} the first law was deduced assuming (baryon) number conservation, and it may have often went unnoticed that actually (\ref{firstlaw}) is the universal form of the law, as will be shown this explicitly later at (\ref{firstlaw2}). In terms of the entropy density $ns$ the first law would read $\diff\rho = T\diff (ns) + \mu\diff n$, consistently with e.g. \cite{Ballesteros:2016kdx,Luongo:2018lgy}.}}
\be \label{firstlaw}
\text{the first law of thermodynamics} : \quad  \diff \rho = nT\diff s + h\diff n\,. 
\ee
The less well-known but fundamental fields discussed in the introduction are the
\bs
\label{fundamental}
\ba
\text{thermasy} & : & \quad \theta\,, \\
\text{chemical momentum} & : & \quad \varphi\,.
\ea
\es
In addition to these thermodynamical quantities, for the consistent Lagrangian formulation of relativistic fluids it is necessary to introduce the \cite{Kijowski:1979dj,Brown:1992kc}
\bs
\ba
\text{Lagrangian coordinates} & : & \quad \alpha^A\,, \\
\text{Lagrange multipliers} & : & \quad \beta_A\,.
\ea
\es
The $\alpha^A(x)$ label the flow lines which pass through each spacetime point $x$. They can be specified on an arbitrary hypersurface since $\beta_A$ imposes their constancy along the fluid flow. 
Finally, we shall state the novel part of our formulation, the interactions. These shall be introduced in terms of the additional vector densities,
\bs
\label{fluxes}
\ba
\text{chemical reaction flux} & : & \quad \mathcal{X}^\mu\,, \\
\text{entropic interaction flux} & : & \quad \mathcal{Y}^\mu\,.
\ea
\es
In this Section, we consider these vector densities as the generic parameterisation of arbitrary interactions. 
Thus, any particular model of interacting hydrodynamics is determined by the given equation of state (\ref{energydensity}) and the given dependence of (\ref{fluxes}) on the fundamental fields. In the following Sections we will then
present concrete realisations of such models.

Finally, we state the generic action functional: 
\be \label{action}
I = \int \diff^4 x\lb -\sqrt{-g}\rho(\lvert\mathcal{J}\rvert/\sqrt{-g},s) + \lp \mX^\mu + {\mJ}^\mu\rp \varphi_{,\mu}   + \lp \mY^\mu + s{\mJ}^\mu\rp\theta_{,\mu} + \mJ^\mu\beta_A\alpha^A{}_{,\mu}  \rb\,.
\ee
We can distinguish the following cases:
\begin{itemize}
\item Non-interacting fluid dynamics is recovered when $\mX^\mu=\mY^\mu=0$. 
\item Non-transverse $\mX^\mu$ results {\it non-conservative} fluid model. The number flux density is not conserved, meaning that effectively, particles can be created and destroyed by the coupling.
\item Non-transverse $\mY^\mu$ results in {\it non-adiabatic} fluid model. Due to non-minimal interactions, the entropy of the fluid is not constant along the fluid flow. (Non-transverse $\mX^\mu$ only affects the specific entropy.) 
\end{itemize}
It is convenient to denote $\mathcal{J}^\mu = \sqrt{-g}J^\mu$, where $J^\mu=nu^\mu$ is a vector, and similarly for the other vector densities $\mX^\mu$ and $\mY^\mu$. By non-transversity of a vector density we thus mean the
non-vanishing of the metric-covariant divergence of the corresponding vector. 
Some possible consequences of the interactions are:
\begin{itemize}
\item If $X^\mu$ or $Y^\mu$ depends on the metric, the couplings may generate heat fluxes and anisotropic stresses \cite{Koivisto:2005mm,Yang:2018ubt,BeltranAlmeida:2019fou,Orjuela-Quintana:2020klr,Arjona:2020kco}.
\item If $X^\mu$ or $Y^\mu$  is a pseudovector, the couplings violate parity \cite{Lue:1998mq,Nishizawa:2018srh,Conroy:2019ibo,Gasparotto:2022uqo,Manton:2024hyc}.
\end{itemize}
The (non-)conservation laws obtained by the variation of $I$ {w.r.t.} the momenta $\varphi$ and $\theta$ are\footnote{Here and in the following we have assumed that the interaction vectors $X^\mu$, $Y^\mu$ do not depend on the fluid variables, but the generalisation with $X(\varphi,\theta)$, $Y(\varphi,\theta)$ is straightforward. We will encounter (effective) generalisations in terms of kinetic terms for the momenta in Section \ref{hyperhydro} and in terms of potential terms for the momenta in Section \ref{cosmo}. However, the interaction vectors should {\it not} depend upon $n$ or $s$ lest the interpretation of $\varphi$ and $\theta$ is modified. }
\bs
\label{nonconservation}
\ba
\Diff_\mu J^\mu & = & -\Diff_\mu X^\mu\,, \label{noncons1} \\
n\dot{s} & = & s\Diff_\mu X^\mu - \Diff_\mu Y^\mu\,,
\ea
\es
where $\Diff_\mu$ is the torsion-free and metric-compatible Levi-Civita covariant derivative. 
The variations {w.r.t.} the particle flux and the entropy per particle result, respectively, in the equations
\bs
\label{fluxeqs}
\ba
h u_\mu + \varphi_{,\mu} + s\theta_{,\mu} + \beta_A\alpha^A{}_{,\mu} & = & 0\,, \label{fluxeq1} \\
\theta_{,\mu}J^\mu & = & \frac{\partial \rho}{\partial s}\,,
\ea
\es
which imply that
\bs
\label{foundation}
\ba
\dot{\theta} & = & T\,, \\
\dot{\varphi} & = & \mu\,. 
\ea
\es
These relations are as foundational as (\ref{firstlaw}) in our formulation and it is crucial that they are not violated despite the possible non-adiabatic or non-conservative properties of the system. 

The variations {w.r.t.}the Lagrange multipliers $\beta_A$ yield
\bs
\begin{equation}
    \dot{\alpha}^A = 0 \,, \label{alpha}
\end{equation}
fixing the fluid four-velocity to be along the flow lines. 
While the Lagrangian coordinate variations usually verify that $\beta_{A}$ are also constants, we instead now have
\begin{equation}
     n\dot{\beta}_A = \Diff_\mu X^\mu\beta_A\,. \label{beta}
\end{equation}
\es
Thus, in the non-conservative case we cannot express the Lagrange multipliers $\beta_A$ as functions of $\alpha^A$. Nevertheless, the vector $\beta_\mu = \beta_A\alpha^A{}_{,\mu}$ provides the orthogonal $u^\mu\beta_\mu=0$ components of the so-called velocity potential representation (\ref{noncons1}) of the 
four-velocity\footnote{The $\beta_A$ can be interpreted as the spatial components of the Taub vector $hu_\mu$ on a hypersurface coordinatised by $\alpha^A$ with $\varphi=\theta=0$ set to zero \cite{Brown:1992kc}.}, consistently with the foundational relations (\ref{foundation}). Therefore, whilst the geometric interpretation of the fluid space hinges on (\ref{alpha}), the right hand side of (\ref{beta}) can in general be nonzero.  

The variations {w.r.t.} to the metric (and the connection, in the generalisation considered in Section \ref{hyperhydro}) determine the gravitational sources. From the metric variations we obtain
\be
\frac{-2}{\sqrt{-g}}\frac{\delta I}{\delta g^{\mu\nu}}  =  T_{\mu\nu} + X_{\mu\nu} + Y_{\mu\nu}\,, 
\ee 
where the fluid energy-momentum tensor and {the interaction energy-momentum tensors} are given by\footnote{The interactions are associated with energy-momentum, except in the (exotic) case that the densities $\mX^\mu$ and $\mY^\mu$ are {metric-independent} “levitons” (if some fields don't couple to the metric at all, they don't  “gravitate” but  “levitate”).}
\bs
\ba
T_{\mu\nu} & = & \rho u_\mu u_\nu + p h_{\mu\nu}\,, \label{matterEMT} \\
X_{\mu\nu} & = &  \lp X^\alpha g_{\mu\nu} - 2\frac{\delta X^\alpha}{\delta g^{\mu\nu}}\rp\varphi_{,\alpha}\,, \\
Y_{\mu\nu} & = &  \lp Y^\alpha g_{\mu\nu} - 2\frac{\delta Y^\alpha}{\delta g^{\mu\nu}}\rp\theta_{,\alpha}\,.
\ea
\es
The result (\ref{matterEMT}) confirms the consistency of the nomenclature in (\ref{energydensity},\ref{pressure}). 
The projection of the metric-covariant divergence of the fluid energy-momentum tensor along the fluid flow gives
\bs
\label{emtcons}
\be
u_\nu\Diff_\mu T^{\mu\nu} = \mu\Diff_\mu X^\mu + T\Diff_\mu Y^\mu\,, \label{energyexh}
\ee
and the orthogonal projection gives
\ba
h_{\alpha\nu}\Diff_\mu T^{\mu\nu} & = & 2n u^\nu\Diff_{[\nu}\lp h u_{\alpha]}\rp  - h^\nu{}_\alpha\frac{\partial \rho}{\partial s} s_{,\nu} 
 =  2n u^\nu\Diff_{[\nu}\lp h u_{\alpha]}\rp - Tn\lp s_{,\alpha} + \dot{s}u_\alpha\rp \nn \\
& = & -\lp Tu_\alpha + \theta_{,\alpha}\rp \lp s\Diff_\mu X^\mu - \Diff_\mu Y^\mu\rp - \beta_A\alpha^A{}_{,\alpha}\Diff_\mu X^\mu\,. \label{momentumexh}
\ea
We note the appearance of the Taub current $v_\mu = h u_\mu$ in the derivation. In the last step we used (\ref{nonconservation}) and (\ref{fluxeqs}). As expected, the fluid energy-momentum tensor is covariantly conserved if the interactions are switched off.

The result for the energy exchange (\ref{energyexh}) completely agrees with our physical intuition. It is given as the sum of the chemical potential times the chemical coupling and the temperature times the entropic coupling. 
The model predicts momentum exchange (\ref{momentumexh}), which consists of a thermodynamical and hydrodynamical contribution. The former is given as the spatial gradient of thermasy times the entropy flow, and the latter involves an acceleration term proportional to $a=u^\mu\partial_\mu\log{\beta_A}$. We may also write the orthogonal projection as 
\be \label{momentumexh2}
h_{\alpha\nu}\Diff_\mu T^{\mu\nu} = -(h_\alpha{}^\mu\theta_{,\mu})\dot{s} - n a \beta_\mu \,,
\ee
\es
to express (\ref{momentumexh}) in a more transparent form\footnote{It is possible to identify $\dot{\beta}_\mu = a\beta_\mu$ in covariant terms when the Levi-Civita connection satisfies $ u^\mu\beta_\alpha \genfrac{\{}{\}}{0pt}{}{\alpha}{\mu\nu} =0$.}. 

A generic perfect fluid is described by five functions. In cosmology, the convention is to identify these as the energy density and the pressure of the fluid, plus the three components of the fluid 3-velocity, which is further decomposed into one scalar and one transverse 3-vector (perturbative) degrees of freedom. In our formulation (\ref{action}), it is natural to consider the five free functions in physical terms as the number density, the entropy and the three spatial components of the Taub current. One may specify initial conditions for these fields at a hypersurface where the torsors $\theta$ and $\varphi$ are set to constants, and then it follows that the spatial components of the Taub current are given directly by (minus) the $\beta_A$, if we choose the $\alpha^A$ as the coordinates of the hypersurface. The five evolution equations for the dynamical fields are then (\ref{nonconservation}) and (\ref{beta}), the thermacy and the chemical potential being determined by the integrals of (\ref{foundation}). 

The sources $\Diff\cdot X$ and $\Diff\cdot Y$ can, in general, involve external fields, render the fluid imperfect and introduce new degrees of freedom. We now proceed to concrete examples of such sources.  

\section{Scalar-fluid models}
\label{scalar-fluid}

Here we consider the fluid interacting with a scalar field $\phi$. Scalar-fluid momentum exhange \cite{Pourtsidou:2013nha,Boehmer:2015kta,Boehmer:2015sha} is a consistent and interesting possibility in cosmology \cite{Koivisto:2015qua,Dutta:2017wfd,Kase:2019veo,Kase:2019mox,Amendola:2020ldb}. {In our new formulation, it follows from the principles of thermodynamics in a generic situation wherein the phase space structure of the $\rho$-fluid may depend on the $\phi$-field ($Y^\mu \neq 0$) or the $\phi$-particles may decay into $\rho$-particles or vice versa ($X^\mu \neq 0$).}   
The main interest in the new models is that they have different observational implications.  

\subsection{Algebraic interactions}

As a simple example, let the interaction vectors be $X^\mu = X^{,\mu}/\Box$ and $Y^\mu = Y^{,\mu}/\Box$, where $X$ and $Y$ are functions of the scalar field and $\Box = \Diff_\mu\Diff^\mu$. Given a canonical Lagrangian for the scalar, we can write the action for the scalar-fluid system (discarding a boundary term) 
\be \label{algebraic}
I = \int \diff^4 x\lb -\sqrt{-g}\rho(n,s) + {\mJ}^\mu\lp \varphi{,_\mu} + s\theta_{,\mu} + \beta_A\alpha^A{}_{,\mu}\rp - \mathcal{X}(\phi)\varphi - \mathcal{Y}(\phi)\theta    - \frac{1}{2}\partial_\mu\phi\partial^\mu\phi - V(\phi)\rb\,.
\ee
The (non-)conservation laws (\ref{nonconservation}) follow directly as
\bs
\ba
\Diff_\mu J^\mu & = & -X\,, \\
\dot{s} & = & sX-Y\,.
\ea
\es
In addition to the fluid $T^{\mu\nu}$, there are scalar field-dependent contributions to the total energy-momentum,
\bs
\ba
T^{(\phi)}_{\mu\nu} & = & \partial_\mu\phi\partial_\nu\phi - g_{\mu\nu}\lb\frac{1}{2}\partial_\alpha\phi\partial^\alpha\phi + V(\phi)\rb\,, \\
X_{\mu\nu} & = & -g_{\mu\nu}X\varphi\,, \\
Y_{\mu\nu} & = & -g_{\mu\nu}Y\theta\,.
\ea
\es
Now the equations of motions are supplemented with the Klein-Gordon equation
\be
\Box\phi - V' = X'\varphi + Y'\theta\,. 
\ee
It is natural to interpret the interaction in terms of an effective potential for the scalar field, $\hat{V} = V + X\varphi + Y\theta$, so that
\be
\hat{T}^{(\phi)}_{\mu\nu}  =  \partial_\mu\phi\partial_\nu\phi - g_{\mu\nu}\lp\frac{1}{2}\partial_\alpha\phi\partial^\alpha\phi + \hat{V}\rp = T^{(\phi)}_{\mu\nu} + X_{\mu\nu} + Y_{\mu\nu}\,,
\ee
and the Klein-Gordon equation reads $\Box \phi = \hat{V}'$. The metric-covariant divergence 
\ba
\Diff^\mu \hat{T}^{(\phi)}_{\mu\nu} & = & \lp\Box\phi - \hat{V}'\rp\phi_{,\nu} - \partial_\nu\lp X\varphi + Y\theta\rp \nn \\
& = & \lp \mu X + TY \rp u_\nu - h^\mu{}_\nu\lp X\varphi_{,\mu} + Y\theta_{,\mu}\rp 
 =  -\Diff^\mu T_{\mu\nu}\,, \label{cons1}
\ea
is compatible with the conservation of the total energy-momentum. In the second step we used the Klein-Gordon equation and made the 1+3 decomposition to compare with (\ref{emtcons}), and in the last step we have used the velocity potential representation from (\ref{fluxeq1}) and noted that now $n a = X$.

A particle creation model using a coupling of the algebraic form was recently considered\footnote{A dynamical system analysis of exponential quintessence with exponential $X$ was performed. Though there $Y=0$, a non-adiabatic interpretation of the model was attempted in terms of heat flow.} in Ref. \cite{Hussain:2024jdt}. 

\subsection{Derivative interactions}

Next, we consider a different structure of the interactions, with the vectors $X^\mu = \partial^{\mu}X(\phi)$,   $Y^\mu = \partial^{\mu} Y(\phi)$ the gradients of some coupling functions $X$ and $Y$. Now the torsor property of the thermodynamical momenta is retained, but the interaction energy-momentum becomes imperfect,
\bs
\ba
X_{\mu\nu} & = & g_{\mu\nu}X'\phi_{,\alpha}\varphi^{,\alpha} - 2X' \phi_{,(\mu}\varphi_{,\nu)}\,, \\
Y_{\mu\nu} & = & g_{\mu\nu}Y'\phi_{,\alpha}\theta^{,\alpha} - 2Y'\phi_{,(\mu}\theta_{,\nu)}\,.
\ea
\es
The effective energy-momentum tensor of the scalar field $\hat{T}^{(\phi)}_{\mu\nu}$ is now described by
\bs
\label{scalarEMT}
\ba
 \text{energy density}:  \quad \hat{\rho}{}^{(\phi)} & = & {-}\frac{1}{2}\lp\partial\phi\rp^2 + V + X'\phi_{,\alpha}\varphi^{,\alpha} + Y'\phi_{,\alpha}\theta^{,\alpha}\,, \\ 
 \text{pressure}:  \quad \hat{p}{}^{(\phi)} & = & {-}\frac{1}{2}\lp\partial\phi\rp^2 - V + X'\phi_{,\alpha}\varphi^{,\alpha} + Y'\phi_{,\alpha}\theta^{,\alpha}\,, \\
\text{{momentum} flux vector}:  \quad \hat{q}_\mu^{(\phi)} & = & \frac{1}{\sqrt{-(\partial\phi)^2}}\lp X'\phi_{,\alpha}\varphi^{,\alpha} + Y'\phi_{,\alpha}\theta^{,\alpha}\rp\phi_{,\mu} + \sqrt{-(\partial\phi)^2}\lp X'\varphi_{,\mu} + Y'\theta_{,\mu}\rp\,. 
\ea
\es
Unlike in the algebraic case, the energetics of the interactions can not be  described simply in terms of an effective potential.  
{If we assume that the fluid is comoving with the scalar field, the above expressions assume a somewhat simpler and more transparent form. Note that the assumption always holds in a Friedmann-Lema{\^i}tre-Robertson-Walker background, and can be adopted as a legitimate gauge choice at the level of cosmological perturbations. We can then write
\bs
\ba
\hat{\rho}{}^{(\phi)} & = & \rho^{(\phi)} + \sqrt{-(\partial\phi)^2}\lp X' \mu + Y' T\rp\,, \\
\hat{p}{}^{(\phi)} & = & p^{(\phi)} + \sqrt{-(\partial\phi)^2}\lp X' \mu + Y' T\rp\,, \\
\hat{q}_\mu^{(\phi)} & = & \sqrt{-(\partial\phi)^2}\lp X'\varphi_{,\mu} + Y'\theta_{,\mu}\rp + \lp X'\mu + Y' T\rp u_\mu\,.  
\ea
\es}
The fluid equations now involve second derivatives of the scalar field,
\bs
\ba
\Diff_\mu J^\mu & = & -X'\Box\phi - X''(\partial\phi)^2\,, \\
n\dot{s} & = & \lp sX'-Y'\rp\Box\phi + \lp sX'' - Y''\rp (\partial\phi)^2\,,
\ea
\es
whereas second derivatives of the thermodynamic potentials appear in the Klein-Gordon equation,
\be
\Box\phi - V' - X'\Box\varphi - Y'\Box\theta = 0\,. 
\ee
It is straightforward to use this Klein-Gordon equation to arrive at the conservation law
\be \label{cons2}
\Diff^\mu \hat{T}^{(\phi)}_{\mu\nu} = -(\Diff_\alpha X^\alpha)\varphi_{,\nu} 
-(\Diff_\alpha Y^\alpha)\theta_{,\nu}\,. 
\ee
Following the same steps as in the previous algebraically coupled case, we verify that both the energy exchange (\ref{energyexh}) and the momentum exchange
(\ref{momentumexh2}) of the fluid is compensated by flow of energy-momentum from $\hat{T}^{(\phi)}_{\mu\nu}$ into conservation of the total energy-momentum. 

It is worth emphasising that the equations~(\ref{cons1}) and~(\ref{cons2}) are the consequence of the diffeomorphism invariance of the theory. Hence, the calculations show that these new models and the formulation in general is internally consistent. This will also be true for the geometric generalisations below.

\section{Hyperhydrodynamical models}
\label{hyperhydro}

An extension of the hydrodynamical action functional with non-minimal couplings of the fluid to the affine connection was called hyperhydrodynamics \cite{Iosifidis:2023kyf}, because such couplings introduce hypermomentum \cite{Hehl:1977gn,Obukhov:1993pt,Brechet:2007cj,Iosifidis:2020gth,Iosifidis:2020upr,Iosifidis:2021nra}. Given a metric, it is customary to characterise a generic affine connection by defining the two other tensors
\bs
\ba
Q_{\alpha\mu\nu} & = & \nabla_\alpha g_{\mu\nu}\,, \\
T^\alpha{}_{\mu\nu} & = & 2\Gamma^\alpha{}_{[\mu\nu]}\,,
\ea
\es
called non-metricity and torsion, respectively. 
Thus, we encode the 40+24=64 
components of an affine connection into the two tensors. 
One quarter of those are contained in the four 1-forms 
\be
Q_\mu = Q_{\mu\alpha}{}{^\alpha}\,, \quad \bar{Q}_\mu = Q^\alpha{}_{\alpha\mu}\,,  \quad T_\mu = T^\alpha{}_{\mu\alpha}\,, \quad \tilde{T}_\mu = \epsilon_{\mu\alpha\beta\gamma}T^{\alpha\beta\gamma}\,,
\ee 
where $\tilde{T}^\mu$ is actually a pseudo-vector. Since the metric is required to construct vectors from the affine connection, hyperhydrodynamical fluids are generically imperfect. 
For generality, we may consider the 4-parameter forms for the two possible couplings,
\bs
\label{hypercouplings}
\ba
X^\mu & \equiv & 
a_1 Q^\mu + a_2\bar{Q}^\mu + a_3 T^\mu + a_4\tilde{T}^\mu \,, \\
Y^\mu & \equiv & 
b_1 Q^\mu + b_2\bar{Q}^\mu + b_3 T^\mu + b_4\tilde{T}^\mu\,,
\ea
\es
where $a_{(i)}, b_{(i)}$ parameters of mass dimension two. 
It is convenient to introduce the shorthand notation 
\be
f^{(i)} = a_{(i)}\varphi + b_{(i)}\theta\,.
\ee
The interaction energy-momentum tensors then assume the form
\bs
\ba
X_{\mu\nu} + Y_{\mu\nu} & = & 
-2\lb Q_{,(\mu}\partial_{,\nu)}  + g_{\mu\nu} \lp \nabla^\alpha\nabla_\alpha  + T^\alpha\partial_\alpha - \bar{Q}^\alpha\partial_\alpha\rp\rb f^{(1)}  \\
& - & \lp 2\nabla_{(\mu} \nabla_{\nu)} + 2T_{(\mu} \partial_{\nu)} + Q_{(\mu} \partial_{\nu)}     - g_{\mu \nu} \bar{Q}^{\alpha} \partial_{\alpha}  \rp f^{(2)}  \\
& - & \lp 2T_{(\mu}\partial_{\nu)} - g_{\mu\nu}T^\alpha\partial_\alpha\rp  f^{(3)}  \\
& - & \lp 2\tilde{T}_{(\mu}\partial_{\nu)} + 4\epsilon_{(\mu|}{}^{\alpha \beta \gamma} T_{|\beta|\nu) \gamma} \partial_{\alpha} - g_{\mu\nu}\tilde{T}^\alpha\partial_\alpha\rp f^{(4)}\,. 
\ea
\es
Now there is also hypermomentum, generated from the variations of the interaction vectors with respect to the independent connection, given by the tensor
\be
Z_\alpha{}^{\mu\nu} \equiv -\frac{1}{\sqrt{-g}}\frac{\delta I}{\delta \Gamma^\alpha{}_{\mu\nu}} = 
 2 \delta_{\alpha}^{\nu} \partial^{\mu} f^{(1)} + \big( g^{\mu \nu} \partial_{\alpha} + \delta^{\mu}_{\alpha} \partial^{\nu}\big) f^{(2)} + 2 \delta_{\alpha}^{[\mu} \partial^{\nu]} f^{(3)} + 2 \epsilon_{\alpha}{}^{\mu \nu \rho} \partial_{\rho} f^{(4)}\,.  
\ee
For a kinematical interpretation, the hypermomentum can be decomposed into 
\bs
\label{hypercomponents}
\ba
 \text{dilation :}  &    \Delta^{\nu} \equiv Z_{\alpha}{}^{\nu\alpha}= \partial^{\nu}\Delta\,, \quad & \Delta  =8 f^{(1)}+2 f^{(2)}-3 f^{(3)}\,, \label{dilation} \\
 \text{spin :} &\  \sigma_{\mu \nu \alpha} \equiv Z_{[\mu| \nu | \alpha]} = g_{\nu [\mu} \partial_{\alpha]} \sigma + 2 \epsilon_{\alpha \mu \nu \rho} \partial^{\rho} \tilde{\sigma} \,, \quad & \sigma = f^{(3)} \,, \ \tilde{\sigma} = f^{(4)} , \label{spin} \\
    \text{shear :} &\ \Sigma_{\mu \nu \alpha} \equiv Z_{(\mu |\nu | \alpha)} - \frac{1}{4} g_{\mu \alpha} \Delta_{\nu} = \big(g_{\nu (\mu} \partial_{\alpha)} - \frac{1}{4} g_{\mu \alpha} \partial_{\nu}\big) \Sigma \,, \quad & \Sigma = 2 f^{(2)} + f^{(3)} \,.
\ea
\es 
Thus, in the present case the couplings (\ref{hypercouplings}) generate only the 4 scalar components of hypermomentum, the dilation, 2 spin scalar and pseudoscalar, and the shear, such that
\be
f^{(1)}=\frac{1}{8}(\Delta -\Sigma + 4\sigma)\,, \quad f^{(2)}=\frac{1}{2}(\Sigma - \sigma)\,,\quad f^{(3)}=  \sigma\,, \quad f^{(4)}= \tilde{\sigma} \,.
\ee
These relations connect thermasy and chemical momentum to spin, dilation and shear. In the more complete theory, we would instead consider the generalised equation of state of the fluid that involves the independent hypermomenta, and couple them to the corresponding irreducible components of the connection. This will of course lead to a generalised first law wherein the hypermomentum appear paired with their respective conjugate variables as indicated in Table \ref{symplectic}. However, our purpose here is not the exhaustive analysis of the generic hyperfluid, but rather to present the necessary proof of the concept. Thus we refrain from introducing new fields into the (\ref{action}) and turn to check the workings of the simplified prescription in the minimal model involving only the scalar pieces of the irreducible decomposition. 

\subsection{Palatini General Relativity}

Let us consider the example of Palatini gravity sourced by a hyperhydrodynamical fluid given by the  interaction fluxes with $a_3=(8a_1 + 2a_2)/3$, $a_4=0$ and $b_{(i)}=b a_{(i)}$ for some constant $b$. The parameter combination is chosen due to the projective invariance of the Palatini action
\cite{Bernal:2016lhq,Iosifidis:2018zwo,Janssen:2018exh,Marzo:2021iok,Sauro:2022hoh} which forces the dilation (\ref{dilation}) to vanish, and the axial spin $\tilde{\sigma}$ in (\ref{spin}) we have excluded for simplicity.
Again it is convenient to define a shorthand for the linear combination of the momentum fields, $f \equiv \varphi + b\theta$. To be explicit, the action is then 
\be \label{hyperaction}
I = \int \diff^4 x\sqrt{-g}\lb \frac{m_P^2}{2}R - \rho(\lvert J \rvert,s) + \lp a_1 Q^\mu + a_2\bar{Q}^\mu + \frac{8a_1+2a_2}{3}T^\mu\rp f_{,\mu} + {J}^\mu\lp \varphi_{,\mu} + s\theta_{,\mu} + \beta_A\alpha^A{}_{,\mu}\rp \rb\,,
\ee
where the pure gravity part of the action is determined by the usual scalar curvature,
\be \label{curvature}
R = g^{\beta\nu}R^\alpha{}_{\beta\alpha\nu} \quad \text{where} \quad
{R}^\alpha_{\phantom{\alpha}\beta\mu\nu}  =  
2\partial_{[\mu} \Gamma^\alpha_{\phantom{\alpha}\nu]\beta}
+ 2\Gamma^\alpha_{\phantom{\alpha}[\mu\lvert\gamma\rvert}\Gamma^\gamma_{\phantom{\lambda}\nu]\beta}\,,
\ee
and $m_P$ is the Planck mass. The equation of motion for the connection gives
\ba \label{pceom}
T^\mu{}_\alpha{}^\nu +Q_\alpha{}^{\mu\nu} +\lp T^\nu+\frac{1}{2}Q^\nu - \bar{Q}^\nu\rp\delta^\mu_\alpha 
 -  \lp T_\alpha+\frac{1}{2}Q_\alpha\rp g^{\mu\nu} & = & 
 -\frac{4}{3m_P^2}\lp a_1+a_2\rp f^{,\mu}\delta^\nu_\alpha
 \nn \\ & + & \frac{2}{3m_P^2}\lp 8a_1 + 5a_2\rp f^{,\nu}\delta^\mu_\alpha
 + \frac{2a_2}{m_P^2}f_{,\alpha}g^{\mu\nu}\,. 
\ea
One of the 3 traces of this equation is identically satisfied due to the projective invariance of our action, the other two are
\begin{subequations}
\label{htraces}
\ba
2 T_\alpha + \frac{3}{2} Q_\alpha -3 \bar{Q}_\alpha & = & \frac{2}{m_P^2}\lp 10 a_1+7a_2\rp f_{,\alpha}\,, \\
 2 T_\alpha + \frac{1}{2}Q_\alpha   + \bar{Q}_\alpha & = & -\frac{2}{m_P^2}\lp 2a_1+5a_2\rp f_{,\alpha}\,. 
\ea
\end{subequations}
The trace equations, as well as the full equation (\ref{pceom}), are solved by\footnote{Connections of a similar form are assumed in the so called vector distortion cosmological models, which consider Riemann-Weyl geometry and its generalisation in terms of a single vector field \cite{BeltranJimenez:2015pnp,BeltranJimenez:2016wxw,Jimenez-Cano:2022sds,TerenteDiaz:2023kgc,Harko:2024fnt}.}
\bs
\label{tracesol}
\ba \label{Tsol}
T^\alpha{}_{\mu\nu} & = & \frac{c}{m_P^2}\delta^{\alpha}_{[\mu}f_{,\nu]}\,, \\
Q_{\alpha\mu\nu} & = & \frac{\lp 4a_1 - 2a_2 + 3c \rp }{3m_P^2}f_{,\alpha}g_{\mu\nu}
- \frac{8\lp a_1 + a_2\rp}{3m_P^2}g_{\alpha(\mu}f_{,\nu)}\,,  \label{Qsol}
\ea
\es
where $c$ is an arbitrary parameter reflecting the projective invariance,  and we chose the parameterisation such that the projective gauge $c=0$ is torsion-free. We then see that the interaction sources in this model become
\bs
\ba
\Diff\cdot X & = & \frac{2}{3m_P^2}\lp 4a_1^2 - 16a_1a_2-11a_2^2\rp \Box\varphi\,, \\ 
\Diff\cdot Y & = & \frac{2b}{3m_P^2}\lp 4a_1^2 - 16a_1a_2-11a_2^2\rp \Box\theta\,. 
\ea
\es
{
The scalar curvature (\ref{curvature}) of the connection with with torsion (\ref{Tsol}) and nonmetricity (\ref{Qsol}) is
\bs
\begin{equation}
    R = R(g) + 2 m_P^{-2}(4 a_1 + a_2) \nabla_{\mu} \nabla^{\mu} f + 2 m_P^{-4} \big( 2 a_1 - a_2 \big)^2 (\partial f)^2 \, ,
\end{equation}
where $R(g)$ is the metric Ricci scalar. Stated in terms of the Levi-Civita d'Alembert operator $\Box = D_{\mu} D^{\mu}$, this takes the form
\begin{align}
    R &= R(g) + 2 m_P^{-2}(4 a_1 + a_2) \Box f - \frac{2}{3} m_P^{-4} \big(4 a_1^2 - 16 a_1 a_2 - 11 a_2^2\big) (\partial f)^2 \, \nonumber \\
    &= R(g) + 2 m_P^{-2}(4 a_1 + a_2) \Box f - m_P^{-2} X^{\mu} f_{, \mu} \, .
\end{align}
\es
Plugging this solution back into the action (\ref{hyperaction}) and discarding dynamically irrelevant boundary terms leads to
\begin{equation}
 I = \int \diff^4 x\sqrt{-g}\lb \frac{m_P^2}{2}R(g) - \rho(\lvert J \rvert,s) 
+ \frac{1}{3m_P^2 }\lp 4a_1^2 - 16a_1 a_2 - 11a_2^2\rp \lp\partial f\rp^2 +  J^\mu\lp \varphi_{,\mu} + s\theta_{,\mu} + \beta_A\alpha^A{}_{,\mu}\rp\rb\,.     
\end{equation}
{Thus, we see that the hyperhydrodynamic interactions can be understood as non-trivial dynamics of the fundamental fields (\ref{fundamental}), since the affine connection can be integrated out} to obtain a theory with non-trivial kinetic terms for the thermasy and the chemical momentum. In general\footnote{ It is not difficult to see that in some simplified cases, in particular, neglecting some of the velocity degrees of freedom residing in the $\alpha^A$ part, the particle flux $\mJ^\mu$ could further be integrated out and, assuming a reasonable equation of state $\rho(n,s)$, some of the cases could be reduced to healthy k-essence or bi-scalar models.}, the nature of the propagating degrees of freedom depends on hyperhydrodynamic coupling parameters $a_1$, $a_2$ and $b$. The Weyl vector coupling contributes generically towards ghost-like kinetic terms, whereas coupling to the other non-metric trace contributes towards healthy kinetic terms. 

\subsection{On general gravity models}

The structure of the hyperhydrodynamic theory is highly sensitive to the choice of the gravitational action. For example, we note that the model we have considered does not exist in the Einstein-Cartan restriction of Palatini gravity (obtained by switching off non-metricity). On the other hand, this suggests new avenues to generalised theories of gravity. Modifications of the Palatini (aka metric-affine) gravity with, for instance, higher curvature invariants quite generically lead to unphysical instabilities   \cite{Percacci:2020ddy,BeltranJimenez:2020sqf,Jimenez-Cano:2022sds,Delhom:2023xdp,Barker:2024ydb} but {as we have demonstrated, modifications of the coupling to matter fields provide an alternative, viable route to new gravitational dynamics}. Models in a different class of modified gravity, formulated in the teleparallel i.e. $R^\alpha{}_{\beta\mu\nu}=0$ geometry, quite generically suffer from strong coupling or ghosts issues \cite{Ong:2013qja,BeltranJimenez:2020fvy,BeltranJimenez:2021auj,Barker:2022kdk,Gomes:2023tur}; the former could perhaps be helped {by exciting the strongly coupled modes by suitable couplings of the affine connection to matter, which arise naturally in the new formulation presented here}. Canonical teleparallel gravity (without those generically pathological modifications) can already be an interesting framework for hyperhydrodynamical applications, since there, by construction, the hypermomentum has a certain conservative structure. Namely, the teleparallel equation of motion for the connection is $({\nabla}_\mu + T_\mu + \frac{1}{2}Q_\mu) Z_\alpha{}^{\mu\nu}=0$ , which sets the $T^{\mu\nu}$ equal to the canonical energy-momentum. We hope to revisit these topics in a future work.



\section{Cosmology}
\label{cosmo}

Finally, we will consider the cosmological implications of the interacting models. We take gravity to be given by the usual General Relativity and adapt the fluid to a flat {Friedmann-Lema{\^i}tre-Robertson-Walker} geometry, so that $u^\mu = \delta^\mu_0$, $h_{0\mu}=0$ and $h_{ij}=a^2(t)\delta_{ij}$, where the latin indices denote the spatial components. The (non-)conservation equations (\ref{nonconservation}) then read
\bs
\label{frwnoncons}
\ba
\dot{n} + 3nH & = & -\Diff\cdot X\,, \\
n\dot{s} -s\Diff\cdot X & = & -\Diff\cdot Y\,,
\ea
\es
and {the energy-momentum conservation laws (\ref{emtcons}) reduce to one non-trivial continuity equation}
\be
\label{frwnoncont}
\dot{\rho} + 3Hnh  =  - \mu\Diff\cdot X - T\Diff\cdot Y\,.
\ee
We readily check that these equations are consistent with the first law of thermodynamics (\ref{firstlaw}), which in the homogeneous and isotropic geometry reduces to $\dot{\rho} = nT\dot{s} - h\dot{n}$. If we consider the energy $E=\rho V$ of the fluid in a given volume $V$, we also have to take into account the differential $p\diff V$ of the hydraulic energy (recall table \ref{symplectic}). We may consider the alternative forms of
\bs
\label{firstlaw2}
\ba 
\text{the first law of thermodynamics} : \quad  \diff E & = & TN\diff s - p\diff V + h \diff N \\
& = & T\diff S - p\diff V + \mu\diff N\,.
\ea
\es
In the first line we introduced $N=nV$, the number of particles in the volume $V$, and in the second line the $S=sN$, the entropy of the fluid. As this is just the rewriting of (\ref{firstlaw}), we confirm its general validity.  In cosmology in particular, since volumes in an expanding (or contracting) universe scale as $V \sim a^3$, we have $\dot{V}/V=3H$ and the above forms of the first law are also easily seen to be consistent with the fluid equations (\ref{frwnoncons}) and (\ref{frwnoncont}).

\subsection{A phase transition of a gas of dark matter-energy particles (emerging from a 3-form)}

{Let us set up the perhaps simplest possible non-trivial scenario}. We take the fluid to be rotationless dust, described by the equation of state $\rho=\mu n$ with $\mu$ a constant, corresponding to zero temperature $T=0$ and zero pressure $p=0$. The cosmological evolution of the thermodynamic momenta,
\bs
\ba
\theta(t) & = & \theta_0\,, \\
\varphi(t) & = & \mu t + \varphi_0\,,
\ea
\es
is then parameterised by three constants. 
{As the aim is now simplicity, we couple this to a cosmological constant instead of a rolling scalar  field}. Thus, we take the algebraic scalar-fluid model (\ref{algebraic}) without a kinetic term  but with the tachyonic mass term $V(\phi) = -m^2\phi^2/2$ and the linear couplings $X(\phi)=-m^3\phi$, $Y(\phi)=-m^3\phi$. 
The evolution of the scalar is then given by $\hat{V}'=0$, so we have $\phi = -m\mu t +\phi_0$, where $\phi_0 = -m(\theta_0+\varphi_0)$. 
The field then contributes the energy 
\be \label{CCterm}
\hat{V} = \frac{1}{2}m^2\lp \phi_0 - m\mu t \rp^2\,,
\ee
and as long as $\phi_0/m \gg  \mu t$, it is indistinguishable from a cosmological constant $\Lambda=m^2\phi_0^2/2$. The density of the interacting dust fluid evolves according to $\dot{\rho} + 3H\rho = \mu m^3\lp \phi_0 - m\mu t\rp$, which in the regime $\phi_0/m \gg  \mu t$ mimics the CDM evolution with an effective source term. If we consider that the mass scale $m$ is of the order of the present Hubble rate and $\phi_0$ is of the order of the Planck mass, the source term will be negligible for a wide range of the parameter $\mu$. By adjusting the dark matter mass $\mu$, the effect of the coupling may become non-negligible between the last scattering and the present, providing a boost which in the dark matter density which could potentially address the Hubble tension. 

\begin{figure}[t]
\includegraphics[width=0.45\textwidth,height=1.0\textheight,keepaspectratio]{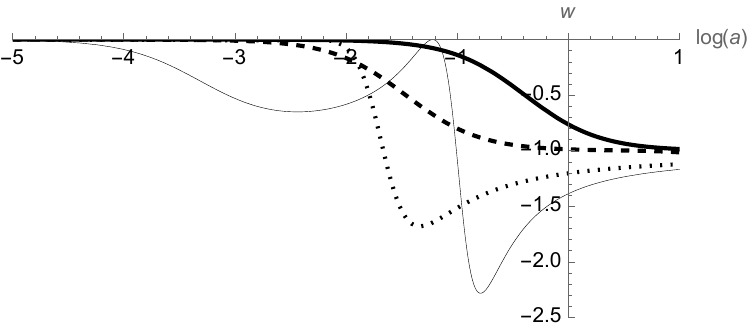}
\includegraphics[width=0.45\textwidth,height=1.0\textheight,keepaspectratio]{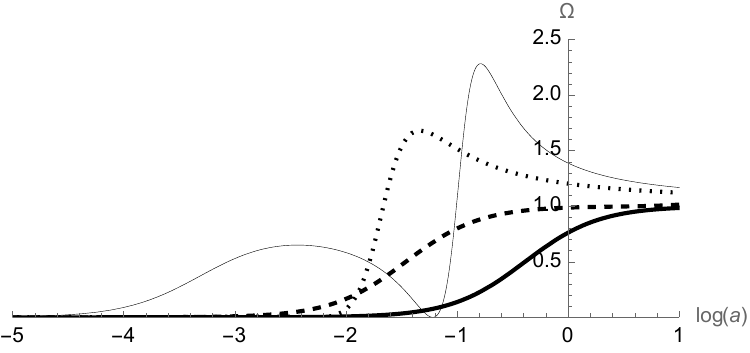}
\caption{\label{3formPlot} Evolution of the cosmological equation of state $w$ (left panel) and the fractional effective dark energy density $\Omega=\hat{V}/3m_P^2H^3$ (right panel) as a function of the e-folding time $\log{a}$ for four parameters and initial conditions chosen such that the four cases exhibit some qualitatively different (not necessarily realistic) features available in the simple model. The thick solid line reproduces the exactly the standard $\Lambda$CDM cosmology. The dashed line shows a faster transition to acceleration. The two other scenarios, plotted with thin solid and thin dotted lines, are characterised by a transition to a phantom-like cosmological equation of state $w<-1$, which is accompanied by effective $\Omega>1$ that implies a negative effective dark matter density, as explained in the text.}
\end{figure}

{The action for this model has the same number of parameters as the $\Lambda$CDM model, the magnitude of $\Lambda$ and CDM now apparently traded for the two mass parameters $\mu$ and $m$}. A physical interpretation of the model is that of self-interacting matter. By integrating out the auxiliary scalar field,
 the action reads, explicitly in terms of the fundamental fields, 
 \be
 I 
  =  \int\diff^4 x\lb -\mu\lvert\mJ\rvert - \frac{1}{2}\sqrt{-g}m^4\lp \varphi + \theta\rp^2 + \mJ^\mu\lp  \varphi_{,\mu} + s\theta_{,\mu}\rp\rb\,.
 \ee
Thus, we have single fluid of particles with mass $\mu$ which are undergoing chemical processes  energy scale is $m$. {In the limit $m\rightarrow0$, the fluid behaves as {\bf pressureless dust} (CDM), and in the limit $\mu\rightarrow 0$ the fluid behaves as vacuum energy ($\Lambda$)}. Since this dust-$\Lambda$ fluid isn't associated with a temperature, we could drop the variables $s$ and $\theta$ without loss of generality in terms of the dynamical solutions. Let us do this, integrate out also the chemical momentum $\varphi$, perform a partial integration and rescale the vector $J^\mu \rightarrow J^\mu/m^2$. This procedure reveals that the thermodynamical fluid model is underpinned by a simple vector field theory,
\be
I = \frac{1}{2}\int \diff^4 x\sqrt{-g}\lb \lp \Diff\cdot J\rp^2 - \mu m^2\sqrt{-J^2}\rb\,.
\ee
{Remarkably, the underlying field theory has only one parameter, $\mu m^2$}. The theory belongs to a class of 3-form models which have been studied previously\footnote{Self-interacting 3-form models have been considered, besides in cosmology, e.g. \cite{Koivisto:2009ew,DeFelice:2012wy,Urban:2013aka,Kumar:2014oka,Barros:2015evi,Wongjun:2016tva,Chakraborty:2021pkp}, in the contexts of stars \cite{Barros:2021jbt}, traversable wormholes \cite{Barros:2018lca,Bouhmadi-Lopez:2021zwt,Tangphati:2023uxt}, regular black holes \cite{Barros:2020ghz,Bouhmadi-Lopez:2020wve}, thick branes \cite{Gordin:2023nsv,Barros:2023nzr}, singularities \cite{Morais:2017vlf,Bouhmadi-Lopez:2018lly}, hamiltonian analysis \cite{Brizuela:2018wmx}, and embedding to metric-affine geometry \cite{Iosifidis:2023pvz}.} \cite{Koivisto:2009sd,Koivisto:2009fb}. This is made apparent by yet one more field redefinition using the Hodge star $\star$ to introduce the 3-form $A = \star J$, whose action then becomes 
\be \label{3form}
I =- \int \diff^4 x\sqrt{-g}\lp \frac{1}{48} F^2 + \frac{\mu m^2}{2\sqrt{6}}\sqrt{A^2}\rp\,,
\ee
where $F_{\alpha\beta\gamma\delta} = 4 \Diff_{[\alpha} A_{\beta\gamma\delta]}$ is the field strength of the 3-form. {It is well-known that the massless 3-form $m^2\mu=0$ provides a $\Lambda$-term \cite{Duff:1980qv,HAWKING1984403,Weinberg:1988cp}, and our result suggests that a potential which is linear in the norm the 3-form can, at least in some dynamical regime, provide also a CDM-term besides the $\Lambda$-term.} In the case of ideal dust the background 3-form components $A_{ijk} = a^3\epsilon_{ijk}m^{-2}n$ would be strictly constant, since they describe the comoving density of the dust. However, the dynamics is more subtle and the evolution of the components $A_{ijk}$ cannot be neglected even in the $\Lambda$CDM-like regime when the Hubble rate is small enough, as seen from the expression for the cosmological equation of state 
\be \label{cEoS}
w \equiv p_A/\rho_A = - \frac{1}{6m^2m_P^2}\lb \frac{\partial \lp a^3 n \rp}{\partial\log{a}}\rb^2\,, 
\ee     
where $m_P$ is again the Planck mass.  The cosmological equation of state vanishes in the early universe and evolves asymptotically to $w \rightarrow -1$ in the future. Figure \ref{3formPlot} illustrates the evolution of $p/\rho$ with four different choices of initial conditions. 
The propagation speed $c_S^2$ of physical fluctuations in the fluid is determined by the potential $V(A^2)$ as \cite{Koivisto:2009fb}
\be
c_S^2 = 1 + 2\frac{V''(A^2)A^2}{V'(A^2)}\,. 
\ee
The sound speed vanishes for a linear potential $V \sim A$. Therefore, the model can reproduce the structure formation of the $\Lambda$CDM cosmology. It is non-trivial that the sound speed remains vanishing even in the dynamical region where the background deviates from $\Lambda$CDM cosmology\footnote{Early proposals of single fluid models for the dark sector \cite{Freese:2002sq,Bento:2002ps} were ruled out by the impact of their non-vanishing sound speed to structure formation \cite{Sandvik:2002jz,Koivisto:2004ne}, and a Lagrangian formulation of a viable unified dark sector remains non-trivial \cite{Li:2009mf,Gao:2009me}.}. 

Finally, we note that the sign of the 3-form potential can be flipped by the dynamics. Then the cosmological equation could become phantom-like $w<-1$, as seen from its expression equivalent to (\ref{cEoS}),
\be
w = -1 + \frac{V}{\rho_A} = -1 + \frac{V}{3m_P^2H^2\Omega_A}\,. 
\ee
The latter formula is in terms of the fraction $\Omega_A$ of the 3-form energy and the total energy of the universe.  
In the fluid picture, the unconventional sign would seem to correspond to negative energy density of the gas of particles, and the action would not remain unbounded from below as the number density of the particles $\lvert n\rvert \rightarrow \infty$. This case could be expected to be pathological, as usual for models with phantom-like $w<-1$. However, validity of the naive stueckelbergisation argument that $V'<0$ implies a ghost  \cite{Koivisto:2009fb} is not obvious for the special case of linear potential. {In addition, whilst for a canonical mass term $V = m_C^2 A^2/2$ one easily establishes the equivalence of the 3-form with a massive scalar field that would be a ghost in the case that the 3-form mass term is tachyonic $m_C^2<0$} \cite{Koivisto:2009fb} but, however, the equivalence with a scalar field breaks down for the linear potential $V  = \mu m^2\sqrt{A^2/24}$. Thus, two standard arguments against $w<-1$ break down for the special case we have arrived at. For curiosity, we plot two examples of cosmological evolution of $w$ in Figure \ref{3formPlot} also in cases such that $w<-1$ occurs. It is interesting that also this (more dubious) model can mimic the $\Lambda$CDM cosmology. Since the result $c_S^2 =0$ holds regardless of the sign of the potential, at least classical instabilities are absent. 

We recall that a dust 3-form appears in the Lorentz gauge theory as an energy source in the effective Einstein equations \cite{Zlosnik:2018qvg}. A massive space arises as a quantum effect in the sense that quantum fluctuations do not satisfy the Hamiltonian constraint of classical General Relativity and may contribute an apparent dust source \cite{Gallagher:2023ghl}. The dust-like 3-form only exists in a symmetry-broken phase of the Lorentz gauge theory, whereas in the model (\ref{3form}) the square root forces an analogous symmetry breaking. It is then tempting to speculate that the model (\ref{3form}) could be an effective description of quantum gravity at cosmological scales, as it seems plausible for the kinetic term of the 3-form to be induced by quantum corrections. Indeed, the 3-form appears linearly in the extension of the Einstein-Cartan action, and there is no reason why the term could not have either sign. 
However, the possible embedding of the model into a fundamental theory of spacetime and gravity, as well as the detailed assessment of the model's compatibility with the data, are outside the scope of the present article. 

\section{Conclusions}
\label{conclu}

We proposed a general action formulation for interactions of relativistic fluids. 
The conceptual framework, 
heuristically summarised in Table \ref{symplectic}, entails strict mathematical requirements for the action functional. For a meaningful hydrodynamical interpretation to exist, we must be able to choose the Lagrangian coordinates $\alpha^A$ on an arbitrary hypersurface, which is tantamount to the statement $\dot{\alpha}^A=0$. However, the conjugate variables $\beta_A$, which can be directly related to the velocity field of the fluid, can be subject to a new dynamical law $\dot{\beta}_A \neq 0$ as the result of the interactions. 
Moreover, a meaningful thermodynamical interpretation requires that the key variables of thermacy and chemical momentum must describe the time integrals of the temperature and the chemical potential, respectively. Again, the conjugates of these variables, the specific entropy and the number density can, in contrast, receive physically motivated corrections to their conservation laws from the new interaction terms in the action. 

The formulation was further developed and its consistency was verified in three classes of scenarios: i) self-interacting fluids, ii) interactions with scalar fields, and iii) hyperhydrodynamical interactions involving metric-affine geometry. New results emerged in each of the three cases. Previously, the idea for hyperhydrodynamics had not yet found its fully satisfactory mathematical expression, as the couplings had been introduced to the four-velocity of the fluid instead of the appropriate thermodynamical momenta. 
On the other hand, scalar field models with non-minimal matter couplings have been already studied extensively and a zoo of viable models can be found in the literature. However, our approach is different with its point of departure in general physical principles, and the resulting models predict new observable signatures that can be constrained with the data and distinguished from other alternative models of the dark sector, in particular by their impact on large scale structures. The systematic study of cosmological perturbation theory of interacting relativistic fluids and the confrontation of the new models with the precision cosmological data are some of the next steps we hope to report in a future publication.



\acknowledgements{We acknowledge useful discussions with Christian Boehmer. 
This work was supported by the Estonian Research Council CoE grant TK202 “Foundations of the Universe” and grant SJD14. 
EJ is supported by the Engineering and Physical Sciences Research Council [EP/W524335/1].
}

\bibliography{hyperhydro}

\end{document}